      [1999/12/01 v1.4c Il Nuovo Cimento]
\documentclass{cimento}

\newcommand{\bi}{\bibitem}
\newcommand{\be}{\begin{eqnarray}}
\newcommand{\ee}{\end{eqnarray}}

\def\be{\begin{equation}}
\def\ee{\end{equation}}
\def\mpl{m_{Pl}}
\def\mn{{\mu\nu}}

\def\-g{\sqrt{-g}}

\usepackage{graphicx}  
\title{Modified gravity. Problems and observational manifestations.}
\author{E.V.~Arbuzova\from{ins:x}
}
 \instlist{\inst{ins:x}Department of Higher Mathematics, University
"Dubna", 141980 Dubna, Russia }

\PACSes{\PACSit{00.00}
\PACSit{98.80.Cq, 95.35.+d, 95.30.Sf}
}
\begin{document}

\maketitle

\begin{abstract}
Some models of modified gravity and their observational manifestations
are considered. It is shown, that gravitating systems with mass
density rising with time evolve to a singular state with infinite curvature scalar. The universe evolution durung the radiation dominated epoch is studied in $R^2$-extended gravity theory. Particle production rate by the oscillating curvature is calculated. 
Possible implications of the model
for cosmological creation of non-thermal dark matter are  discussed.
\end{abstract}

\section{Introduction}
Discovering of the cosmic antigravity based on the accumulated astronomical data, such as observation of the large scale structure of the universe,
measurements of the angular fluctuations of the cosmic microwave background radiation, determination of the universe age
(for a review see~\cite{ref:cosm-prmtr}), and especially discovery of the dimming of distant Supernovae~\cite{ref:Nobel_2011},
is the most attractive event in cosmology of the last quarter of century. It was established and unambiguously proved that the universe expansion is accelerated, but the driving force behind this accelerated expansion in still unknown.

Among possible explanations, the most popular is probably
the assumption of a new (unknown) form of cosmological energy density with large negative pressure, $ P < -\rho/3$, the so-called dark energy, for a review see e.g.~\cite{ref:DE_Peebles_Ratra}. 

A competing mechanism to describe the accelerated expansion is represented by gravity modifications at small curvature, the so-called $f(R)$-gravity theories, as suggested in ref.~\cite{ref:grav-mdf}. In these theories the standard
Einstein-Hilbert Lagrangian density, proportional to the scalar curvature $R$, is replaced by a function
$f(R)$, so the usual action of General Relativity acquires an additional term:
\be
 \label{e.A1}
S = -\frac{m_{Pl}^2}{16\pi} \int d^4 x \sqrt{-g}\,f(R)+S_m=
-\frac{m_{Pl}^2}{16\pi} \int d^4 x \sqrt{-g}\,\left[R+F(R)\right]+S_m\, ,
\ee
where $m_{Pl}= 1. 2 2\cdot 10^{19}$ GeV is the Planck mass and $S_m$ is the matter action.

The original version of such models~\cite{ref:grav-mdf} suffers from a strong instability in presence of gravitating 
bodies~\cite{ref:DolgKaw} and because of that more complicated functions $ F(R) $ have been 
proposed~\cite{ref:Starob, ref:HuSaw, ref:ApplBatt, ref:Noj-Odin-2007}, which are free from the mentioned exponential 
instability. 

Though free of instability~\cite{ref:DolgKaw}, the models proposed in~\cite{ref:Starob, ref:HuSaw, ref:ApplBatt} possess
another troublesome feature, namely in a cosmological situation they should evolve from a singular
state in the past~\cite{ref:appl-bat-08}. Moreover, it was found in refs.~\cite{ref:frolov, ref:Arb_Dolgov}
that in presence of matter, a singularity may arise in the future if the matter density rises with time; such future singularity is unavoidable, regardless of the initial conditions, and is reached in a time which is much shorter than the cosmological one.  

\section{Explosive phenomena in modified gravity}

In paper~\cite{ref:Arb_Dolgov} the model of modified gravity with $F(R)$ function suggested in ref.~\cite{ref:Starob} was considered: 
\be
F(R)= \lambda R_0 \,\left[ {\left(1+ \frac{R^2}{R_0^2}\right)^{-n}} - 1 \right]\,.
\label{e.F-AAS}
\ee
Here constant $\lambda $ is chosen to be positive to produce an
accelerated cosmological expansion, $n $ is a positive integer, and 
$R_0$ is a constant with dimension of the curvature scalar.  
The latter is assumed to be of the order of
the present day average curvature of the universe, i.e. 
$R_0 \sim 1/t^2_U$, where $t_U \approx 4\cdot 10^{17}$ sec is the universe age.

The corresponding equations of motion have the form
\be
\left( 1 + F'\right) R_{\mu\nu} -\frac{1}{2}\left( R + F\right)g_{\mu\nu}
+ \left( g_{\mu\nu} D_\alpha D^\alpha - D_\mu D_\nu \right) F'  = 
\frac{8\pi T^{(m)}_{\mu\nu}}{m_{Pl}^2}\,,
\label{e.eq-of-mot}
\ee 
where $F' = dF/dR$, $D_\mu$ is the covariant derivative,  and 
$T^{(m)}_{\mu\nu}$ is the energy-momentum tensor of matter.

By taking trace over $\mu $ and $\nu $ in eq.~(\ref{e.eq-of-mot}) we obtain the equation
of motion which contains only the curvature scalar $R$  and the trace
of the energy-momentum tensor of matter: 
\be
3 D^2 F' -R + R F' - 2F = T\,,
\label{e.D2-R}
\ee
where $T = 8\pi T_\mu^\mu /m_{Pl}^2$.

We analize evolution of $R$ in massive objects with time  varying mass density ,
$\rho_m \gg \rho_c $.  The
cosmological energy density at the present time is 
$\rho_c \approx 10^{-29}\,{\rm g/cm}^3$, while matter density of, say,
a dust cloud in a galaxy could be about $\rho_m \sim 10^{-24}{\rm g/cm}^3$.   
Since the magnitude of the curvature scalar is proportional to the
mass density of a nonrelativistic system, we find $R \gg R_0 $. In
this limit: 
\be
F(R) \approx -\lambda R_0 \left[ 1 -\left(\frac{R_0}{R}\right)^{2n} \right] \,.
\label{e.F-large-R}
\ee

The equation of motion is very much simplified if we introduce the new notation  
\be
w= - F' =  2n\lambda \left(\frac{R_0}{R}\right)^{2n+1} \,.
\label{e.F'-of-R}
\ee

Evolution of $w$ is governed by a simple equation of unharmonic oscillator:
\be
(\partial^2_t - \Delta) w  + U'(w) = 0\,.
\label{e.eq-for-w}
\ee
Potential $U(w)$ is equal to:
\be
U(w) = \frac{1}{3}\left( T - 2\lambda R_0\right) w + 
\frac{R_0}{3} \left[ \frac{q^\nu}{2n\nu} w^{2n\nu}+ \left(q^\nu
    +\frac{2\lambda}{q^{2n\nu} } \right) \,\frac{w^{1+2n\nu}}{1+2n\nu}\right]\,,
\label{e.U-of-w}
\ee
where $\nu = 1/(2n+1)$, $q= 2n\lambda$, and in
eq. (\ref{e.eq-for-w}) $U'(w)=dU/dw$. 

Notice that infinite $R$ corresponds to $w=-F' =0$, so if 
$F'$ reaches zero, it would mean that $R$ becomes infinitly large.

Potential $U$ would depend upon time, if the mass
density of the object under scrutiny changes with time, $T=T(t)$.
If only the dominant terms are retained 
and if the space derivatives are neglected, equations
(\ref{e.eq-for-w}), (\ref{e.U-of-w}) simplifie to:
\be
z'' - z^{-\nu} + (1+\kappa \tau) = 0\,.
\label{e.eq-for-z}
\ee 
Here we introduced dimensionless quantities  
\be
t = \gamma \tau,\,\,\,
\gamma^2 = \frac{3q}{(-R_0)} \left(-\frac{R_0}{T_0}\right)^{2(n+1)}\,, \, \, 
 w = \beta z\,,\ \
\beta = \gamma^2T_0/3 = q \left(-\frac{R_0}{T_0}\right)^{2n+1}\,.
\label{e.dimless}
\ee

\begin{figure}
\begin{center}
\includegraphics[scale=0.6]{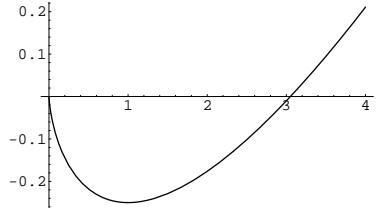}
      \caption{ Potential $
U(z)=z (1+\kappa \tau)  - z^{1-\nu }/(1-\nu ),\ \nu=\frac{1}{5}, \tau=0. 
$ }
   \label{f-1}
\end{center}
\end{figure}

\begin{figure}
\begin{center}
\hspace{-0.7cm}
    \includegraphics[scale=0.7]{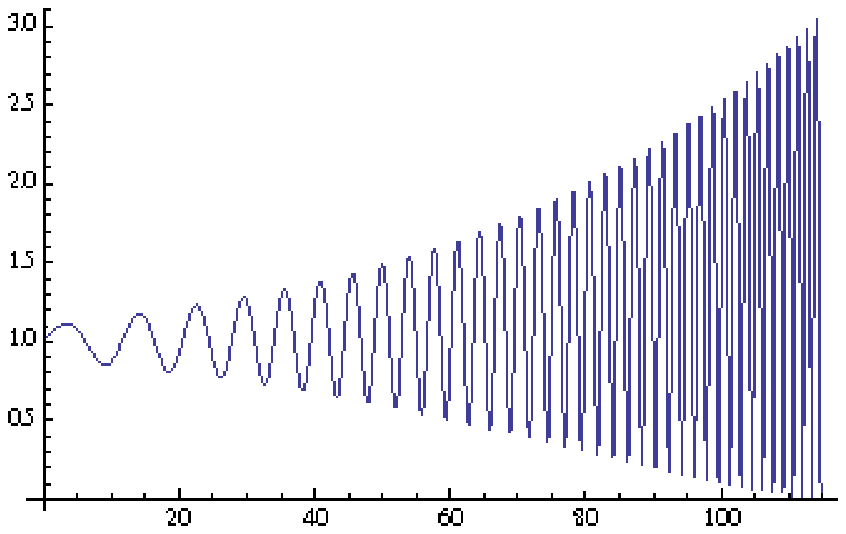} 
\hspace{.2cm}
   \includegraphics[scale=0.7]{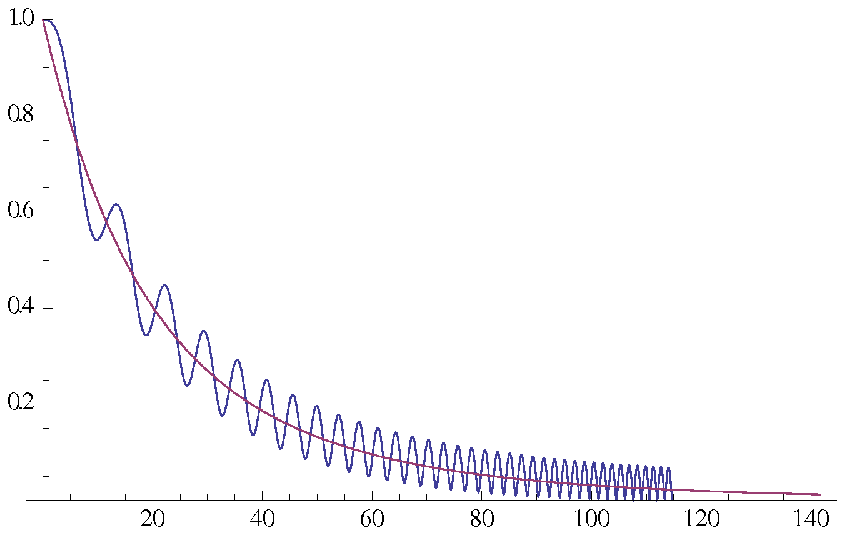}
       \caption{Ratio $z(\tau)/z_{min}(\tau)$ (left)
and  functions $z(\tau)$ and $z_{min}(\tau)$ (right) for $n=2$, $\kappa =0.01$, $\rho_m/\rho_c = 10^5$. The initial conditions: $z(0) =1$ and $z'(0) =0$. }
   \label{f-12}
\end{center}
\end{figure}

Minimum of the potential $U(z)$ (Fig.~\ref{f-1}) sits at $z_{min} = (1+\kappa \tau)^{-1/\nu}$. When the mass density rises, the minimum moves towards zero and becomes less deep.
If at the process of "lifting" of the potential  $z(\tau)$ happens to be
at $U>0$
it would overjump potential which is equal to zero at $z=0$.  In other words, 
$z(\tau )$ would reach zero, which corresponds to infinite $R$, and so
the singularity can be reached in finite time (see Fig.~\ref{f-12}).

The simplest way to avoid singularity is to introduce $R^2$-term into the
gravitational action:
\be
\delta F(R) = -R^2/ 6m^2\,,
\label{e.delta-F}
\ee
where $m$ is a constant parameter with dimension of mass. 

In the homogeneous case and in the limit
of large ratio $R/R_0$ equation of motion for $R$ is modified as:
\be 
\left[ 1-\frac{R^{2n+2}}{6\lambda n(2n+1) R_0^{2n+1} m^2 }\right]\,\ddot R
- (2n+2) \,\frac{\dot R^2}{R}
-\frac{R^{2n+2} (R+T)}{6\lambda n (2n+1) R_0^{2n+1}} = 0 \,.
\label{e.eq-for-R-mdf}
\ee

With dimensionless curvature and time
\be 
y = -\frac{R}{T_0}\,,\,\,\,\,
\tau_1 =  t \left[-\frac{T_0^{2n+2}} { 6\lambda n (2n+1) R_0^{2n+1}}
\right]^{1/2} 
\label{e.y-x}
\ee
equation for $R$ is transformed into:
\be 
\left(1 + g y^{2n+2} \right) y'' 
- 2(n+1)\,\frac{(y')^2}{y} +y^{2n+2} \left[y - (1 + \kappa_1 \tau_1)\right] = 0 \,,
\label{e.y-2prime}
\ee
where prime means derivative with respect to $\tau_1$.

\begin{figure}
\begin{center}
\hspace{-1.0cm}
    \includegraphics[scale=0.7]{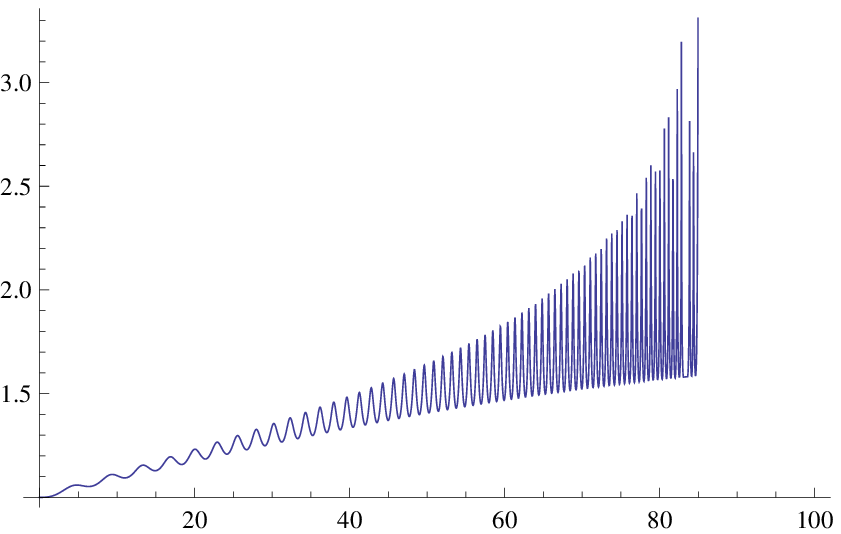} \hspace{.3cm}
 \includegraphics[scale=0.7]{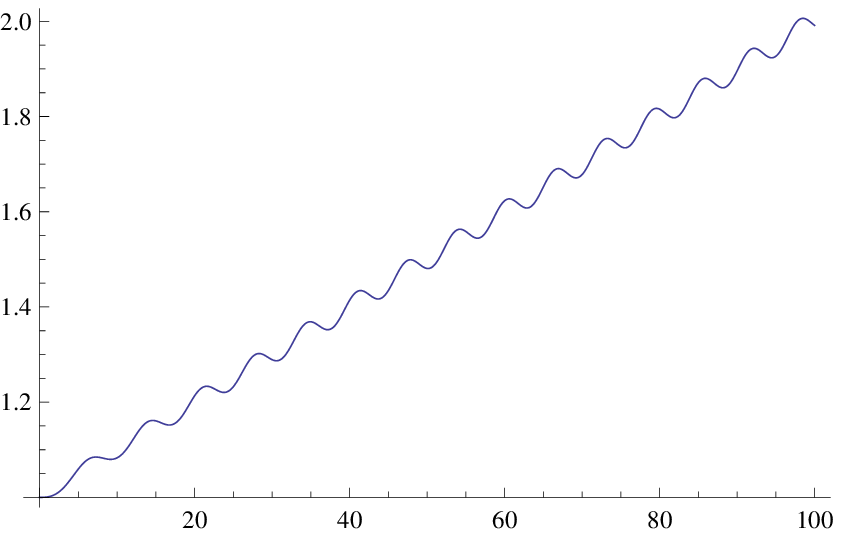}
       \caption{Numerical solutions of eq.~(\ref{e.y-2prime}) for $n=3,\  \kappa_1 =0.01$,  
$y(\tau_{in})=1+\kappa_1 \tau_{in}, \ \ y'(\tau_{in})=0$. Left panel: $g=0$. Right panel: $g=1$. }
   \label{f-g0g1}
\end{center}
\end{figure}

We introduced here the new parameter, $g$, which can prevent from the approach to infinity and is equal to:
\be
g = -\frac{T_0^{2n+2}}{6\lambda n (2n+1) m^2 R_0^{2n+1}}>0 \,.
\label{e.g}
\ee 
For very large $m$, or small $g$,
when the second term in the coefficient of the second derivatives in eqs.~(\ref{e.eq-for-R-mdf}) 
and (\ref{e.y-2prime}) can be neglected,  
numerical solution demonstrates that $R$ would reach
infinity  in finite time in accordance with the results presented above (see Fig.~\ref{f-g0g1}, left panel).
Nonzero $g$ would terminate the unbounded rise of $R$. To avoid too
large deviation of $R$ from the usual gravity coefficient $g$ should be larger than
or of the order of unity. In the right panel of Fig.~\ref{f-g0g1} it is clearly seen, that for $g=1$ the amplitude of oscillations remains constant whereas the average value of $R$ increases with time.
As follows from eq. (\ref{e.y-2prime}), the frequency of small oscillations
of $y$ around $y_0 = 1+\kappa_1 \tau_1$ in dimensionless time  $\tau_1$ is
\be
\omega_\tau^2  = \frac{1}{g}\,\frac{g y_0^{2n+2}}{1+g y_0^{2n+2}} \leq \frac{1}{g}
\label{e.omega-tau}
\ee
It means that in physical time the frequency would be 
\be
\omega \sim \frac{1}{t_U} \left(\frac{T_0}{R_0}\right)^{n+1}
\frac{y_0^{n+1}}{\sqrt{1+g y_0^{2n+2}}} \leq m\,.
\label{e.omega}
\ee
In particular, for $n= 5$ and for a galactic gas cloud with $T_0/R_0 = 10^5$, 
the oscillation frequency would be 
$10^{12}\,\, {\rm Hz} \approx 10^{-3}$ eV.  Higher density objects
e.g. those with $\rho =1\,\, {\rm g/cm}^3$ would oscillate with much higher
frequency, saturating bound (\ref{e.omega}), i.e. $\omega \sim m$.
All kind of  particles with masses smaller than $m$
might be created by such oscillating field.




\section{Cosmological evolution and particle production in $R^2$ gravity} 

In the present secton we study the cosmological evolution of the Universe in a theory with
only an additional $R^2$ term in the action, neglecting other terms which have been introduced
to generate the accelerated expansion in the contemporary universe \cite{ref:Arb_Dolg_Rev}. The
impact of such terms is negligible in the limit of sufficiently large curvature, $|R|\gg |R_0|$, where
$R_0$ is the cosmological curvature at the present time.

In other words, we study here the cosmological evolution
of the early and not so early universe in the model with the action:
\be
S = -\frac{m_{Pl}^2}{16\pi} \int d^4 x \sqrt{-g} \left(R-\frac{R^2}{6m^2}\right)+S_m\,.
\label{e.A-R2}
\ee

The modified Einstein equations for this theory read:
\be
 R_{\mn} - \frac{1}{2}g_{\mn} R -
 \frac{1}{3m^2}\left(R_{\mn}-\frac{1}{4}R g_{\mn}+
g_{\mu\nu} D_\alpha D^\alpha - D_\mu D_\nu \right)R 
 =\frac{8\pi}{\mpl^2}T_\mn\,. \label{e.field_eqs}
\ee

Expressing the curvature scalar $R$ through the Hubble parameter $H = \dot a/a$ as $R=-6\dot H-12H^2$, we get the time-time component of eq.~(\ref{e.field_eqs}):  
\be
\ddot H+3H\dot H - \frac{\dot H^2}{2H}+\frac{m^2 H}{2} = \frac{4\pi m^2}{3\mpl^2 H}\rho\,,
\label{e.timetime}
\ee
where over-dots denote derivative with respect to physical time $t$.

Taking trace of eq.~(\ref{e.field_eqs}) yields:
\be
\ddot R + 3H\dot R+m^2\left(R+\frac{8\pi}{\mpl^2}T^\mu_\mu\right)=0\,.
\label{e.trace}
\ee

In what follows, we study the cosmological evolution in the $R^2$-theory assuming rather
general initial conditions for $R$ and $H$ and dominance of relativistic matter
with the following equation for the matter content:
\begin{equation}\label{e.rho_evol}
\dot \rho + 4H\rho =0\,.
\end{equation}

It is convenient to rewrite the equations in terms of the dimensionless quantities \mbox{$\tau=H_0\,t$}, $h=H/H_0$, $r=R/H_0^2$,
$y=8\pi\rho/(3\mpl^2 H_0^2)$, and $\omega=m/H_0$,where $H_0$ is the value of the Hubble parameter at some initial time $t_0$. Thus the following
 system of equations for dimensionless Hubble parameter is obtained:
\begin{eqnletter}\label{e.hubble_evolution}
h'' + 3h h' - \frac{h'^2}{2h}+\frac{\omega^2}{2}\frac{h^2-y}{h}=0\,, \\
y' + 4hy = 0\,.
\end{eqnletter}

First we assume that the deviations from General Relativity (GR) are small and expand
$h=1/(2\tau) +h_1$ and $y=1/(4\tau^2)+y_1$, assuming that $h_1/h \ll
1$ and $y_1/y \ll 1$, and solve the linearized system of equations. 

\begin{figure}[!h]
\centering
\includegraphics[width=.4\textwidth]{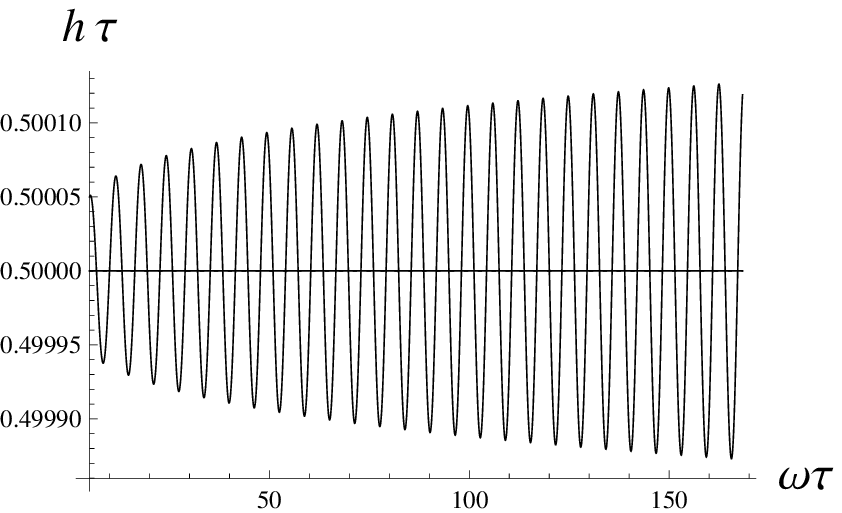}
\includegraphics[width=.4\textwidth]{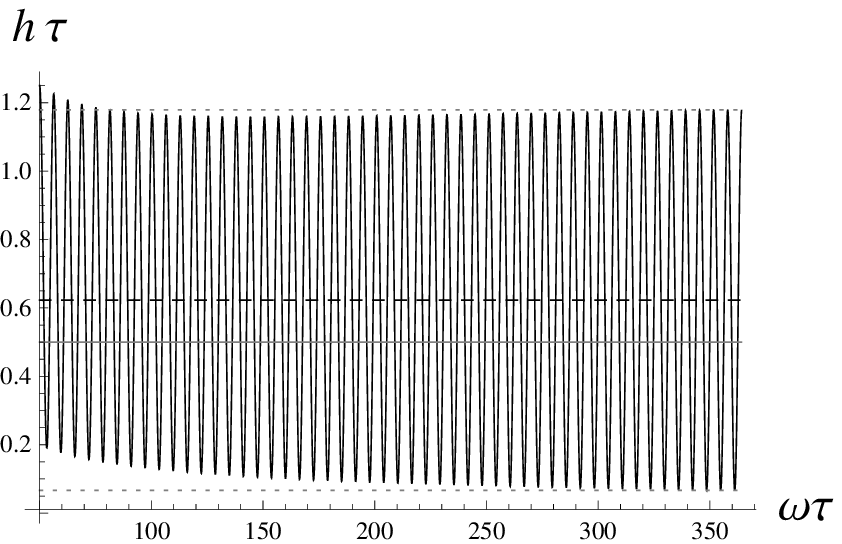}
 \label{f-lin}
\caption{Numerical solution 
of eqs.~(\ref{e.hubble_evolution}). Left panel: linear regime 
 with $\delta h_0=10^{-4}$,  
$\delta h'_0=0$, $y_0=1$,
$\omega=10$. Right panel:  high frequency limit 
with $\delta h_0=1.5$,  
$\delta h'_0=0$, $y_0=0$,
$\omega=100$. Initial conditions are different from GR, the central value $h\tau =0.6$ is shifted from GR value 0.5.}
\label{fig:htau}
\end{figure}

The complete asymptotic solution for $h$ has the form:
\begin{equation}\label{e.sol_h_linear}
h(\tau)\simeq \frac{1}{2\tau}+\frac{c_1\sin(\omega\tau+\varphi)}{\tau^{3/4}}\,.
\end{equation}
The Hubble parameter oscillates around GR value, $h_0\sim 1/(2\tau)$ with rising amplitude,
$h_1/h_0\sim \tau^{1/4}$, and
for sufficiently large $\tau$ the second term would start to dominate and the linear approximation would no longer hold.
Using trancated Fourier expansion it is possible to obtain approximate
analytical solutions of the full non-linear system in the high-frequency limit
$\omega\tau\gg 1$. The same results are found numerically for the initial conditions $h_0=1+\delta h_0\,, h'_0=-2+\delta h'_0\,\ y_0=1+\delta y_0$ 
(see Fig.~\ref{fig:htau}).

Gravitational particle production may non-trivially affect the
solutions of the above equations. Below we consider particle
production by the external oscillating gravitational field and present the equation of motion for the
evolution of $R$ with the account of the back-reaction from particle production. This leads to an exponential
damping of the oscillating part of $R$, while the non-oscillating "Friedmann" part remains practically undisturbed. 
The particle production influx into the cosmological plasma is estimated in the case of a massless, minimally-coupled to gravity scalar field with the action:
\begin{equation}\label{phi_action}
 S_\phi=\frac{1}{2}\int d^4x\,\sqrt{-g}\,g^{\mu\nu}\partial_\mu\phi\,\partial_\nu\phi\,.
\end{equation}

It terms of  the conformally rescaled field, $\chi\equiv a(t)\phi$, and conformal time $\eta$, such that $a\,d\eta=dt$, we can rewrite the equations of motion as:
\begin{eqnletter}\label{e.sys_conformal_R_chi}
&&R''+2\frac{a'}{a}R'+m^2a^2R=8\pi\frac{m^2}{\mpl^2}\frac{1}{a^2}\left[\chi'^2-(\vec\nabla\chi)^2+\frac{a'^2}{a^2}\chi^2-\frac{a'}{a}(\chi\chi'+\chi'\chi)\right], \label{e.R-diprime}\\
&&R=-6a''/a^3\,, \label{e.R}\\
&& \chi''-\Delta\chi+\frac{1}{6}\,a^2R\,\chi=0\,,
\label{e.chi-diprime}
\end{eqnletter}

We derive a closed equation for $R$ taking the average value of the $\chi$-dependent quantum operators
in the r.h.s. of eq.~(\ref{e.R-diprime}) over vacuum, in presence of an external classical gravitational field $R$ following the procedure described in ref.~\cite{ref:Dolgov_Hansen}, where such
equation was obtained in one-loop approximation.

The dominant contribution of particle production is  given by equation: 
\begin{equation}\label{e.R_with_back_reaction_approx}
 \ddot R+3H\dot R+m^2R\simeq 
\frac{1}{12\pi}\frac{m^2}{\mpl^2}\int_{t_0}^tdt'\,\frac{\ddot R(t')}{t-t'}\,.
\end{equation}
This equation is linear in $R$ and naturally non-local in time since the impact of particle production depends upon all
the history of the evolution of the system. 

Using again the procedure of truncated Fourier expansion 
including the back-reaction effects in the form of equation (\ref{e.R_with_back_reaction_approx}), we obtaine the decay rate:
\be
\Gamma_R =
\frac{m^3}{48m_{Pl}^2}\,.
\label{e.Gamma-R}
\ee
This result is in agreement with ref.~\cite{Vilenkin_1985}. The characteristic decay time of the oscillating curvature is
\be
\tau_R = \frac{1}{2\Gamma_R} =  \frac{24 m^2_{Pl}}{m^3} \simeq 2
\left(\frac{10^5\tx{ GeV}}{m}\right)^3 \tx{ seconds}\,.
\label{e.tau-R}
\ee

The contribution of the produced particles into the total cosmological energy density reaches its maximum value at approximately this time.

The influx of energetic protons and antiprotons
could have an impact on BBN. Thus this would either allow to obtain some bounds on $m$ or even to improve the agreement between the
theoretical predictions for BBN and the measurements of the primordial abundances
of light nuclei.

The oscillating curvature might be a source of dark matter in the form of heavy supersymmetric (SUSY)
particles. Since the expected light SUSY particles have not yet been discovered at LHC, to some people supersymmetry somewhat
lost its attractiveness. The contribution of the stable lightest SUSY particle into the cosmological energy
is proportional to
\be
\Omega \sim{ m^2_{SUSY} }/m_{Pl}
\label{Omega-SUSY}
\ee
and for $m_{SUSY} $ in the range $100-1000$ GeV the cosmological fraction of these particles would be
of order unity. It is exactly what is necessary for dark matter. However, it excludes thermally produced LSP's if they are much
heavier. If LSP's came from the decay of $R$ and their mass is larger than the ``mass'' of $R$, i.e. $m$,
the LSP production could be sufficiently suppressed to make a reasonable contribution to dark matter.

 In contemporary astronomical objects oscillation frequency
could vary from $m$ down to very low frequency. The oscillations may
produce radiation from high energy cosmic rays down to radio waves.

\acknowledgments

I thank my coauthors, A.D. Dolgov and L. Reverberi, for cooperation.
I am especially grateful to A.D. Dolgov for help and fruitful discussions
during the entire period of collaboration.
This work was supported by the Grant of the Government of Russian Federation, No. 11.G34.31.0047.


\begin{thebibliography}{0}
\bibitem{ref:cosm-prmtr}
\BY{Nakamura~K. et al. (Particle Data Group)} \IN{J. Phys. G}{37}{2010}{075021}.

\bibitem{ref:Nobel_2011} \BY{Riess~A.G. et al.}
\IN{Astron. J.}{116}{1998}{1009-1038}; \IN{Astrophys. J.}{607}{2004}{665-687};\\
\BY{Perlmutter~S. et al.} \IN{Nature}{391}{1998}{51}; \IN{Astrophys. J.}{517}{1999}{565-586};\\
\BY{Schmidt~B.P. et al.} \IN{Astrophys. J.}{507}{1998}{46}.

\bibitem{ref:DE_Peebles_Ratra} \BY{Peebles~P.J.E., Bharat Ratra}\IN{Rev. Mod. Phys.} {75}{2003}{559-606};\\
\BY{Copeland~E.J., Sami~M. \atque Tsujikawa~S.} \IN{Int. J. Mod. Phys.}{D15} {2006} {1753-1936}.

\bibitem{ref:grav-mdf}
\BY{Capozziello~S., Carloni~S. \atque Troisi~A.}
\IN{Recent Res. Dev. Astron. Astrophys.}{1}{2003}{625},
\BY{ Carroll~S.M., Duvvuri~V., Trodden~M., Turner~M.S.}
\IN{Phys. Rev.} {D 70}{2004} {043528}

\bibitem{ref:DolgKaw}
\BY{ Dolgov~A.D., Kawasaki~M.} \IN{Phys. Lett. B} {573}{2003}{1}.

\bibitem{ref:Starob}
\BY{ Starobinsky~A.A.} \IN{JETP Lett.}{86}{2007}{157}.

\bibitem{ref:HuSaw}
\BY{Hu~W., Sawicki~I.} \IN{Phys. Rev. D} {76} {2007} {064004}.

\bibitem{ref:ApplBatt}
\BY{Appleby~A., Battye~R.} \IN{Phys. Lett. B} {654}{2007}{7}.

\bi{ref:Noj-Odin-2007}
\BY{Nojiri~S., Odintsov~S.D.} \IN{Phys. Lett.B }{657}{2007}{238}.


\bi{ref:appl-bat-08}
\BY{Appleby~S.A., Battye~R.A.} \IN{JCAP}{0805}{2008}{019}.

 \bi{ref:frolov}
\BY{Frolov~A.V.} \IN{Phys. Rev. Lett.}{101}{2008}{061103};\\
\BY{Thongkool~I., Sami~M., Gannouji~R., Jhingan~S.} \IN{Phys. Rev. D}{80}{2009} {043523};\\
\BY{I. Thongkool~I., M. Sami~M., Rai Choudhury~S.} \IN{Phys. Rev. D}{80}{2009}{127501}.

 \bi{ref:Arb_Dolgov}
\BY{Arbuzova~E.V., Dolgov~A.D.} \IN{Phys. Lett.}{B}{700}{2011}{289}.

\bi{ref:Arb_Dolg_Rev}
\BY{Arbuzova~E.V., Dolgov~A.D., Reverberi~L.} \IN{JCAP}{02}{2012}{049} 

\bi{ref:Dolgov_Hansen}
\BY{Dolgov~A.D., Hansen~S.H.} \IN{Nucl.Phys.}{B548}{1999}{408-426}.

\bibitem{Vilenkin_1985} 
\BY{Vilenkin~A.} \IN{Phys. Rev.}{D32}{1985}{2511}.

\end{thebibliography}
\end{document}